\documentclass[12pt,a4paper]{article}

\usepackage{authblk}
\usepackage{graphicx}
\usepackage[font={small}]{caption}
\usepackage{pdfpages}

\def\TiO2{TiO$_2$}
\def\WO3{WO$_3$} 	
\def\Al2O3{Al$_2$O$_3$} 
\def\V2O5{V$_2$O$_5$}

\DeclareMathSymbol{\mh}{\mathord}{operators}{`\-}

\begin{document}



# Monomeric, Two-Dimensionally Ordered WO$_3$ Clusters on Anatase TiO$_2$(101)

Tao Xu,[1] Kræn C. Adamsen,[1] Hanne Falsig,[2] Søren B. Rasmussen,[2] Zheshen Li,[3]

Stefan Wendt,[1,*] and Jeppe V. Lauritsen[1,*]

[1]*Interdisciplinary Nanoscience Center (iNANO), Aarhus University, Gustav Wieds Vej 14, 8000 Aarhus C, Denmark*

[2]*Haldor Topsøe A/S, Haldor Topsøe Allé 1, DK-2800 Kongens Lyngby, Denmark*

[3]*ISA, Department of Physics & Astronomy, Aarhus University, Ny Munkegade 120, 8000 Aarhus C, Denmark*

*Corresponding author: jvang@inano.au.dk; swendt@phys.au.dk

We combined scanning tunneling microscopy (STM) and X-ray photoelectron spectroscopy (XPS) experiments with density functional theory (DFT) calculations to study dispersed tungsta clusters on anatase TiO$_2$(101). Following two different preparation methods, we found that monomeric WO$_3$ species are the most stable configuration rather than WO$_3$ trimers, (WO$_3$)$_3$. The WO$_3$ monomers form tetrahedral WO$_4$ structures on anatase TiO$_2$(101), with one W–O bond and two W–O–Ti linkages per WO$_3$ monomer. Locally, the WO$_3$ monomers form well-ordered (2×1) structures. The discovered geometric structure of WO$_3$ on anatase TiO$_2$(101) opens up numerous opportunities for fundamental studies addressing tungsta and accurate structure-activity studies of WO$_3$ / TiO$_2$ model catalysts.

**PACS numbers:** 68.47.Gh, 82.65.+r , 36.40.Mr, 61.46.Bc

1

WO$_3$ supported on anatase TiO$_2$ (a-TiO$_2$) has numerous applications, for example in the selective catalytic reduction (SCR) of NO$_x$ [1–4], photo-catalysis [5, 6], and photoelectrical water-splitting [7, 8]. The widespread applications of this material motivated studies to elucidate the geometric structure of WO$_3$-monolayers (WO$_3$-ML's) on a-TiO$_2$, which is essential for gaining insight into structure-activity relationships. For example, how WO$_3$ acts as a promoter in SCR catalysis is currently intensely debated [3, 4]. Thus, the knowledge of the geometric structure may elucidate what effects are predominating in the promotion effect, which can help designing catalytic materials with improved performance.

Using TiO$_2$ powders, UV-vis, Raman and XRD studies suggest that WO$_3$ exists as monomeric and polymeric surface species in the sub-ML range, and a crystalline WO$_3$ phase forms above the first ML [9, 10]. A NEXAFS study on a polycrystalline (TiO$_2$ P25) carrier suggests that WO$_3$ exists as tetrahedral WO$_4$ or pentahedral WO$_5$ units, that can be interlinked by W–O–W bonds, forming 2-D islands at high coverage [11]. Despite the great value of such experimental data, the various polymorphs (often intermixed), defects, and facets, that are all present in common TiO$_2$ powders, compromise the conclusions regarding the atomic structure of supported WO$_3$-ML's.

Previous studies of WO$_3$ supported on metal single crystals such as Cu(110) [12, 13], Pd(100) [14] and Pt(111) [15] revealed that well-ordered tungsta overlayers can be formed. Regarding oxide surface supports, Bondarchuk *et al*. studied WO$_3$ supported on rutile TiO$_2$(110) [r-TiO$_2$(110)] by STM [16, 17]. These authors found monodispersed tungsta species and assigned them to trimeric WO$_3$ [(WO$_3$)$_3$] species [16, 17]. On the basis of this assignment, Zhu *et al*. [18] performed DFT calculations addressing (WO$_3$)$_3$ adsorption on r-TiO$_2$(110) and proposed two isoenergetic configurations to correspond to the tungsta species observed by STM. However, Zhu *et al*. [18] did not consider any structures of dissociated (WO$_3$)$_3$ species. Such dissociated (WO$_3$)$_3$ species were reported to exist on Pt-supported ML FeO(111) films, where protrusions appear as "equilateral triangles" that may arise

from three arranged $WO_3$ monomers [19]. Thus far, there is no study addressing $WO_3$-ML's supported on a-$TiO_2$(101), even though a-$TiO_2$ powders are used as support in most applications [20].

In this Letter, we report on the structure of tungsta on a-$TiO_2$(101) in the sub-ML range. The (101) facet was chosen, since this is the most stable facet of a-$TiO_2$, representing the dominant surface exposed by a-$TiO_2$ nanoparticles that are used in most applications. In addition, the a-$TiO_2$(101) surface is well-studied by numerous techniques [21–26]. We used two different methods for $WO_3$ deposition: (i) reactive deposition of metallic W in $O_2$ and (ii) $WO_3$ powder sublimation. We found that monomeric $WO_3$ are formed on a-$TiO_2$(101) at room temperature (RT), irrespective of the preparation method. The $WO_3$ monomers form tetrahedral $WO_4$ structures on the a-$TiO_2$(101) surface, with one W–O(2f) bond and two W–O–Ti(5f) linkages per $WO_3$ monomer.

The experimental and computational details addressing the UHV apparatus, sample preparations, $WO_3$ deposition parameters, STM scanning parameters, XPS experiments and DFT calculations are described in the Supporting Information [27]. Briefly, the STM experiments were conducted in a UHV chamber with a base pressure in the low $10^{-11}$ Torr range [33] equipped with a homebuilt, variable-temperature Aarhus STM [34]. Electrochemically etched tungsten tips were used. If not stated otherwise, the STM scanning was conducted at RT.

Fig. 1a,b shows high-resolution STM images of a clean a-$TiO_2$(101) surface that is characterized by trapezoidal shaped terraces of various sizes [22–26]. The a-$TiO_2$(101) surface is unreconstructed and has a sawtooth-like structure with ridges of two-fold coordinated, bridging O atoms [O(2f)] along the [010] direction, see Fig. 1c. In addition, there are five-fold coordinated Ti(5f) and six-fold coordinated Ti(6f) surface atoms. On the very surface, there are no O vacancies [23]. One ML is defined as the number of Ti(5f) sites, $5.16 \times 10^{14}$ cm$^{-2}$. Before the preparation of $WO_3$ / a-$TiO_2$(101) samples, we checked by STM whether the a-$TiO_2$(101) surface is indeed clean and well-ordered.

3

We prepared WO$_3$ / a-TiO$_2$(101) samples in the sub-ML to ML range with the sample held at RT. Using the reactive deposition method, it is known that predominantly monomeric WO$_3$ is deposited onto the surface, because tungsten evaporates in high vacuum essentially as atoms like many other metals [35]. Indeed, the mass spectrum of a heated W filament in O$_2$ is dominated by the signals of WO$_3$ monomers and its compounds [36]. In contrast, WO$_3$ powder sublimation is known to generate cyclic (WO$_3$)$_3$ species (WO$_3$ trimers) in the gas phase [12, 37]. Our XPS measurements (see Fig. S1) conducted at the MatLine at the ASTRID2 synchrotron facility revealed that both preparation methods lead to supported tungsta species that are predominantly in the +6 oxidation state (W 4f$_{7/2}$ at ~35.7 eV).

Figures 2a–c show STM images of reactive-deposited WO$_3$ species at sub-ML coverage. The main feature on this sample are the bright protrusions (denoted as "M features") that are randomly distributed on the a-TiO$_2$(101) substrate. The M features are very homogeneous in their lateral STM sizes (~6.2 Å) and STM heights (~1.5 Å). In addition to isolated M features, there are examples of M features that are in close vicinity to each other, forming chain-like structures (denoted as "C features"). Figures 2b,c show zoom-in STM images of the M features (b) and C features (c), respectively. The C features are oriented either along [$\bar{1}$11] or [11$\bar{1}$] directions of the a-TiO$_2$(101) substrate. The superimposed blue dots in Figs. 2b,c indicate the STM protrusions of the bare a-TiO$_2$(101) substrate that originate from the O(2f) / Ti(5f) pairs [23, 26]. It can be seen that the M features are confined in the smallest surface triangles formed by the a-TiO$_2$(101) surface protrusions, and the C features can be thought to be composed of several neighboring M features aligned along [$\bar{1}$11] and [11$\bar{1}$] directions. However, along the [010] direction, the nearest M neighbors are separated by one lattice unit. Thus, the C features exhibit a (2×1) superstructure on a-TiO$_2$(101). STM images of WO$_3$ / a-TiO$_2$(101) samples with higher WO$_3$ coverages are presented in Fig. S2.

Figures 2d,e show typical STM images of powder sublimated WO$_3$ species on a-TiO$_2$(101) at sub-ML coverage. Before STM scanning at RT, the sample was annealed at 500 K for 5 minutes to obtain a well-ordered surface. The smallest new features appearing on the a-TiO$_2$(101) surface are the M features, identical to those observed following reactive deposition. C features also appeared on the surface, again aligned along the [$\bar{1}$11] and [11$\bar{1}$] directions. Whereas both preparation methods led to M and C features, there is a clear difference in the distribution of these features on the a-TiO$_2$(101) surface. Following WO$_3$ powder sublimation, the M and C features appear mostly as multiples of three (3M, 6M, etc.). This is particular obvious in Fig. 2e. To assess the distribution of M and C features quantitatively, we performed a stochastic analysis to obtain the nearest-neighbor distance distributions corresponding to the two preparation methods of WO$_3$ / a-TiO$_2$(101), see Fig. S3. This analysis confirmed that the M features are homogeneously distributed in case of reactive deposition and that groups of M features in multiples of three are predominant for WO$_3$ / a-TiO$_2$(101) samples prepared by WO$_3$ powder sublimation.

The data presented thus far point to an assignment of the M features to monomeric WO$_3$ clusters. The main rationale for this assignment is as follows: Since the sizes, shapes and adsorption sites of the smallest tungsta species are identical following both preparation methods (see Fig. 2) and their oxidation state is +6 (see Fig. S1), the predominant M features on both types of WO$_3$ / a-TiO$_2$(101) samples must have identical structures. Considering that the sublimation of WO$_3$ powder leads to cyclic (WO$_3$)$_3$ in the gas phase [12, 37], the WO$_3$ trimers probably dissociate into three monomeric WO$_3$ units, either upon landing on the surface or annealing of the sample at 500 K. This would explain why the M features appear preferentially in groups of three (3M) in case of powder sublimation, but not in case of reactive deposition of W in O$_2$, where the creation of dispersed monomeric WO$_3$ is expected [35, 36].

To test this assignment, we deposited WO$_3$ powder onto a-TiO$_2$(101) at low temperatures, see Fig. 3. Supposed the temperature is sufficiently low, we expect to find intact cyclic (WO$_3$)$_3$ species. Figure 3a shows an STM image of WO$_3$ / a-TiO$_2$(101) when the WO$_3$ powder was sublimated onto a-TiO$_2$(101) at 165 K. Following this preparation, the tungsta species (marked in Fig. 3a by white circles) appear as triangular-shaped, fuzzy protrusions with edge lengths of ~12.0 Å, and there are no M features. With a view on the preparation conditions, and because of their triangular shape, see Fig. 3b, the protrusions seen in Fig. 3a likely arise from intact (WO$_3$)$_3$ species. The triangular shape fits well to the symmetry of gas phase cyclic (WO$_3$)$_3$ species.

Figure 3c shows the STM image of the same WO$_3$ / a-TiO$_2$(101) sample as in Fig. 3a but after brief annealing at 500 K. Now the M features are the smallest tungsta species and they are grouped in clusters of three, just as found in Fig. 2e (see also Fig. 3d). Comparing the numbers of distinguishable tungsta clusters in Fig. 3a and Fig. 3c, the number is about three times higher after annealing at 500 K. Thus, annealing of the WO$_3$ / a-TiO$_2$(101) sample at 500 K leads to the dissociation of (WO$_3$)$_3$ species [38]. This experiment provides strong evidence that the M features arise from WO$_3$ monomers.

Motivated by these experimental observations and previous works [39, 40], we performed a set of DFT calculations, the details of which are described in [27]. Briefly, the DFT calculations were performed using the Grid based Projector Augmented Wave (GPAW) package through the ASE package. The a-TiO$_2$(101) surface was modelled with a (3 × 1) super cell and a slab consisting of three TiO$_2$ trilayers. Within LCAO (linear combination of atomic orbitals) mode, a genetic algorithm (GA) was used to obtain the most stable structures of the (WO$_3$)$_x$ species on the a-TiO$_2$(101) surface, which were subsequently optimized in the more accurate grid-based mode with h = 0.18. The generalized gradient approximation was used with the BEEF-vdW functional. Using this approach, we compared the thermodynamic stability of the most stable monomeric WO$_3$, dimeric W$_2$O$_6$, and

(WO$_3$)$_3$ clusters on a-TiO$_2$(101), see Figs. 4, S4−S6 and Tables 1 and S1. All the $E_{W_xO_{3x},\ stability}$ energies of the structural models are given per WO$_3$ unit and referenced to the identified most stable adsorbed WO$_3$ monomer, that is, to the A1 structure, see Figs. 4a and S4a. How $E_{W_xO_{3x},\ stability}$ was calculated is described in [27]. A positive $E_{W_xO_{3x},\ stability}$ value means that the considered tungsta cluster is less stable than the A1 structure.

The A1 structure is a WO$_3$ monomer that is adsorbed on the a-TiO$_2$(101) surface via two W–O–Ti(5f) linkages and one W–O(2f) bond (see Figs. 4a and S4a). The W atom is located in a tetrahedral environment formed by the neighboring O atoms. This structure is ~0.7 eV more stable than the identified second most stable WO$_3$ monomer, the A1.1 structure depicted in Fig. S4b. Notice that numerous configurations were tested in our approach, because we used a GA.

The computed most stable WO$_3$ dimer and trimer structures are presented in Figs. 4b and 4c,d. There is no good match between the WO$_3$ dimer and trimer structures and the a-TiO$_2$(101) substrate. As a result, the most stable WO$_3$ dimer (A2) is only bound via three bonds to the substrate – similar as the A1 structure. Following similar arguments, it is plausible that an intact cyclic (WO$_3$)$_3$ cluster (configuration A3) with four bonds to the substrate is less stable than a configuration consisting of a W$_3$O$_8$ species and an O$_{ads}$ (A3.1).

Comparing the A1 structure to the most stable WO$_3$ dimers and trimers ($E_{W_xO_{3x},\ stability}$), we find that the most stable WO$_3$ monomer is more stable than the identified most stable dimer and trimer structures (Tables 1 and S1). Accordingly, a model consisting of disintegrated three WO$_3$ monomers is preferred over all other possible models, consisting of larger (WO$_3$)$_x$ units. That is, the dissociation of (WO$_3$)$_3$ species into three WO$_3$ monomers is anticipated, in agreement with our STM analysis.

With the identified most stable WO$_3$ monomer structure (Fig. 4a) it becomes clear why no direct neighboring WO$_3$ monomers are observed along the [010] direction by STM, i.e., why a (2×1)

superstructure forms at high $WO_3$ coverages. Two neighboring Ti(5f) sites along the [010] direction are needed to bind one $WO_3$ monomer, and each Ti(5f) site can be occupied only by one monomer. Accordingly, the $WO_3$ monomers are separated by at least one lattice spacing along the [010] direction (see Figs. 2c and S7). However, along the $[\bar{1}11]$ and $[11\bar{1}]$ directions, there is no such competition for Ti(5f) sites. Hence, the $WO_3$ monomers form a (2×1) superstructure on a-$TiO_2$(101), and a full $WO_3$-ML is reached at 0.5 ML.

Having assigned the M features with great certainty to $WO_3$ monomers, we finally address the lateral sizes of $WO_3$ monomers and cyclic $WO_3$ trimers in the STM images. A cyclic $WO_3$ trimer in the gas phase has a diameter of ~5.3 Å [36]. The larger appearance of $(WO_3)_3$ species in our STM images (~12 Å edge length, see Fig. 3a) likely arises from topological and electronic tip effects [41, 42]. Such "tip broadening" is quite common. For example, the size of supported metal clusters is often overestimated by STM [43]. Note further that a low-temperature STM study of intact $(WO_3)_3$ species on Cu(110) also reported an edge length of ~12 Å [12]. In this light, it is plausible that $WO_3$ monomers appear in STM with lateral sizes of ~6.2 Å and heights of ~1.5 Å.

Given that the STM appearance of $WO_3$ monomers on a-$TiO_2$(101) found here is very similar to that of tungsta species supported on r-$TiO_2$(110), see the studies by Bondarchuk *et al*. and Kim *et al*. [16, 17], we found that the assignment of tungsta species to intact cyclic $(WO_3)_3$ species put forward in these works should be revisited. We thus prepared additionally $WO_3$ / r-$TiO_2$(110) samples by $WO_3$ powder deposition, see Fig. S8. We observed groups of M features (3M) on $WO_3$ / r-$TiO_2$(110), very similar as in case of $WO_3$ / a-$TiO_2$(101) [28]. This is clear indication that r-$TiO_2$(110)-supported tungsta species are likewise $WO_3$ monomers rather than intact cyclic $(WO_3)_3$ species.

The identification of monomeric $WO_3$ clusters on a-$TiO_2$(101) opens numerous possibilities for future studies. Firstly, the physics and chemistry occurring on $WO_3$ / $TiO_2$ model catalysts can now be investigated and understood. We stress that deep fundamental insights are exclusively possible

with the knowledge of the exact atomic structure. Opportunely, the chemistry on the $WO_3$ / a-$TiO_2$(101) model catalytic system is well-accessible because of its high thermal stability. Because anatase is used in most applications and the (101) face is predominant on a-$TiO_2$ nanoparticles, this model system is very close to real heterogeneous catalysts, allowing one to draw conclusions addressing real catalysts. For example, that $WO_3$ monomers line up in 2D rows along the [$\bar{1}$11] and [11$\bar{1}$] directions on a-$TiO_2$(101), as uncovered here, may lead to a paradigm shift in the understanding of the $WO_3$ promotion effect on $V_2O_5$ in SCR catalysis [44]. Secondly, the $WO_3$ / a-$TiO_2$(101) systems may also allow us to study surface processes on tungsta itself. Notice that tungsta is an oxide material with numerous very interesting properties, but a model system mimicking the processes on tungsta surfaces is lacking so far. Finally, we anticipate that the structural characteristics found here for $WO_3$ / a-$TiO_2$(101) maybe transferable to other metal oxide species on oxide supports, such as $MO_3$ / $TiO_2$, $VO_x$ / $TiO_2$ and $CrO_3$ / $TiO_2$. As such, $WO_3$ / a-$TiO_2$(101) has the potential to became a prototypical system for metal oxide species on oxide supports and mixed oxides.

In conclusion, monomeric $WO_3$ clusters form readily on a-$TiO_2$(101), irrespective of the preparation method. For $WO_3$ powder sublimation, this means that cyclic $(WO_3)_3$ species dissociate on the surface. However, the preparation method influences the distribution of the $WO_3$ monomers: In case of reactive deposition of metallic W in $O_2$, the $WO_3$ monomers are homogeneously distributed across the surface. In case of $WO_3$ powder sublimation, the $WO_3$ monomers appear preferentially in groups of three, because they are formed through $(WO_3)_3$ dissociation. The $WO_3$ monomers form locally (2×1) superstructures on a-$TiO_2$(101). Our DFT calculations confirm that monomeric $WO_3$ species are more stable than intact $(WO_3)_3$ species. In addition, we revealed the detailed structure of the $WO_3$ monomers, which explains the (2×1) superstructure observed experimentally. Utilizing high-resolution STM, the described methodology allowed us to clearly distinguish whether the monodispersed tungsta species are intact or dissociated $(WO_3)_3$ species.

**Table 1.** Relative stabilities of tungsta monomers, dimers and trimers on a-TiO$_2$(101). The $E_{W_xO_{3x},\ stability}$ energies are given per WO$_3$ and with respect to the most stable adsorbed WO$_3$* monomer (* represents an adsorption site). Structural details and the total DFT energies are provided in [27].

| Structural model | $E_{W_xO_{3x},\ stability}$ (eV) |
|---|---|
| WO$_3$*(A1) | 0 |
| WO$_3$*(A1.1) | 0.7 |
| W$_2$O$_6$*(A2) | 0.47 |
| W$_2$O$_6$*(A2.1) | 0.53 |
| W$_3$O$_9$*(A3) | 0.42 |
| W$_3$O$_8$* + O*(A3.1) | 0.37 |

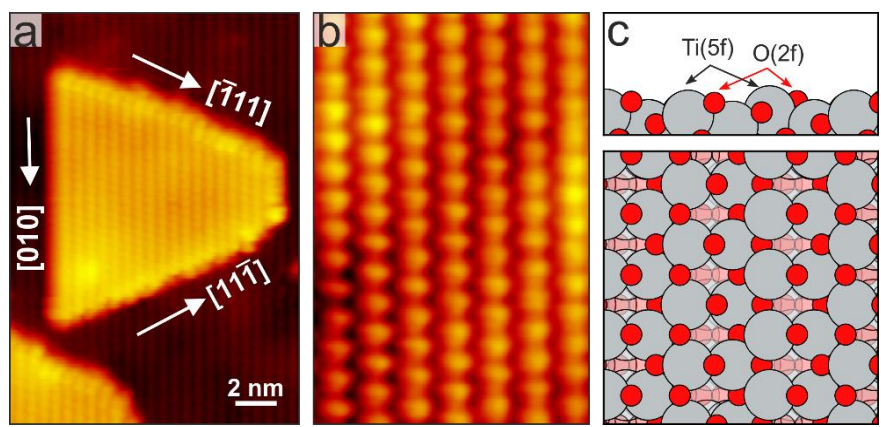

FIG 1. (a,b) High-resolution STM images [a) 140 Å × 200 Å, b) 40 Å × 60 Å] of the bare a-TiO$_2$(101) surface. (c) Ball model of the a-TiO$_2$(101) surface: Gray balls: Ti atoms; red balls: surface O atoms. Pink balls: subsurface O atoms. Under-coordinated surface O(2f) and surface Ti(5f) atoms are indicated by arrows.

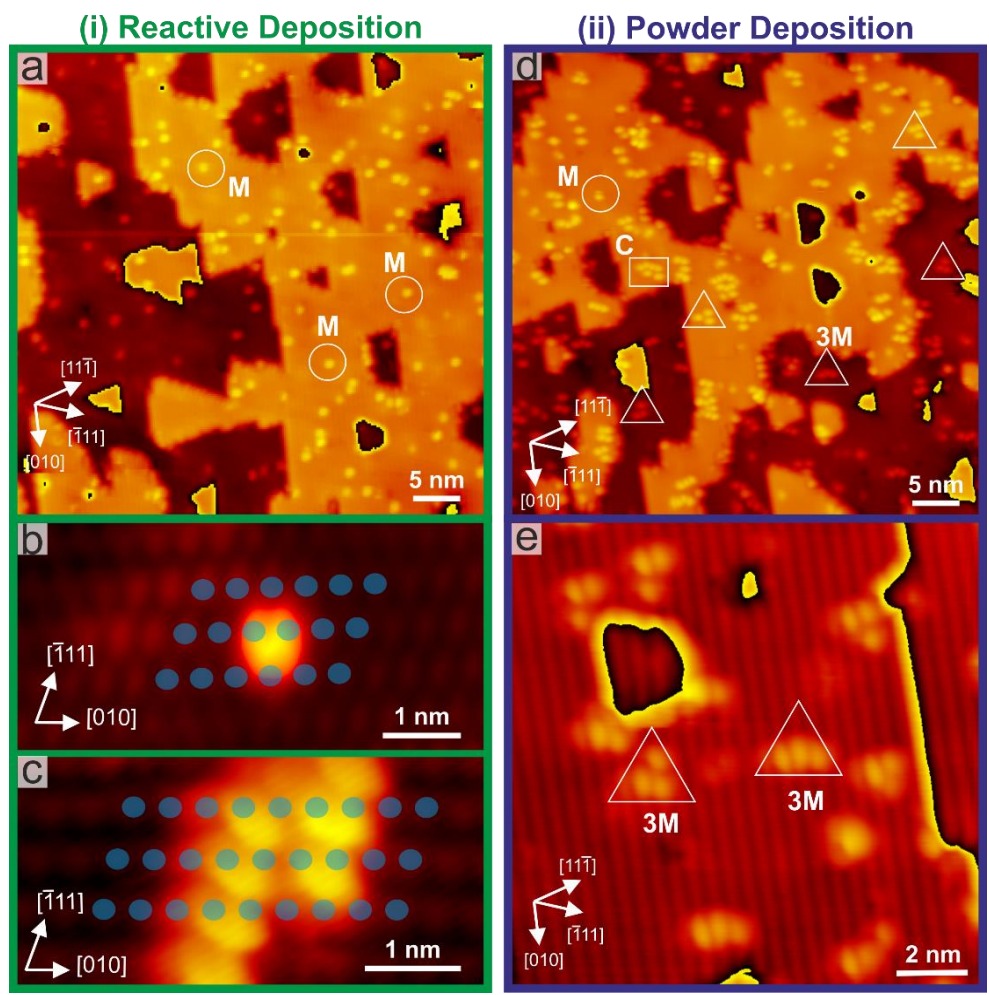

FIG 2. (a−c) STM images [a) 500 Å × 500 Å, b) 60 Å × 30 Å, c) 50 Å × 25 Å] of reactive-deposited WO$_3$ on a-TiO$_2$(101) at low coverage. White circles indicate some of the M features. Zoom-in STM images of the M (b) and C features (c). Blue dots indicate STM protrusions of bare a-TiO$_2$(101). (d, e) STM images [d) 500 Å × 500 Å, e) 150 Å × 150 Å] of powder sublimated WO$_3$ on a-TiO$_2$(101). 3M features (white triangles) and C features (white rectangles) are indicated.

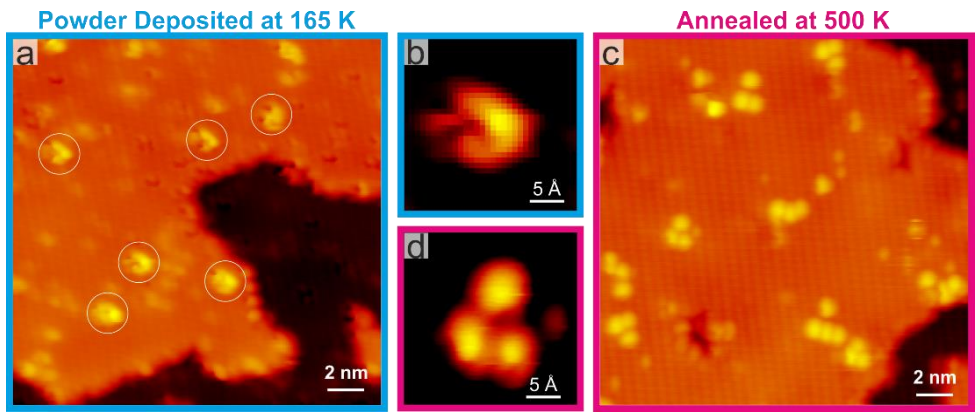

FIG. 3. (a) STM image (200 Å × 200 Å) of $WO_3$ / a-$TiO_2$(101) prepared by $WO_3$ powder deposition at 165 K. (b) Zoom-in STM image (25 Å × 25 Å) of the protrusions marked by white circles in (a) that are assigned to $(WO_3)_3$ species that are possibly still intact. STM images in (a, b) were acquired at 120 K. (c) STM image (200 Å × 200 Å) of $WO_3$ / a-$TiO_2$(101) acquired after the sample corresponding to Fig. 3(a,b) was annealed at 500 K for 5 minutes. (d) Zoom-in STM image (25 Å × 25 Å) of 3M features as seen in (c) that are assigned to three adjacent $WO_3$ monomers. STM images in (c, d) were acquired at RT.

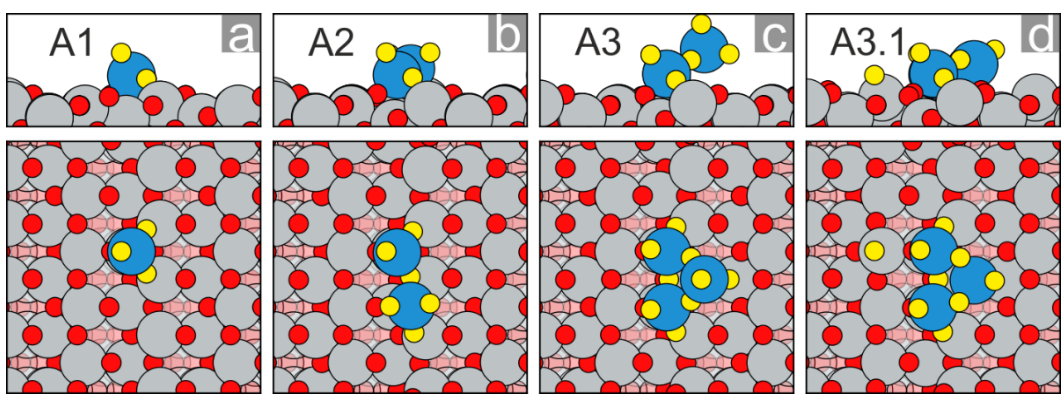

FIG. 4. Computed most stable adsorption structures of (a) $WO_3$ (A1), (b) $W_2O_6$ (A2), (c) $W_3O_9$ (A3) and (d) $W_3O_8 + O_{ads}$ (A3.1) on a-$TiO_2$(101) in side (upper panel) and top (lower panel) views: Gray balls: Ti atoms; red balls: surface O atoms; blue balls: W atoms; Yellow balls: O atoms of $WO_3$ and adsorbed O ($O_{ads}$); Pink balls: subsurface O atoms.

17



\title{\textbf{Supporting Information}\\
 \vspace{5mm}
Monomeric, Two-Dimensionally Ordered \WO3 Clusters on Anatase \TiO2(101)}
\author{Tao Xu, Kr\ae n C. Adamsen, Hanne Falsig, Søren B. Rasmussen, Zheshen Li, Stefan Wendt* and Jeppe V. Lauritsen*}
\date{}
\maketitle

\begin{center}
\textbf{Contents}
\end{center}

\begin{flushleft}
1. a-\TiO2(101) surface preparation
\end{flushleft}
2. \WO3 deposition methods\\
\\
3. XPS of \WO3/a-\TiO2(101)\\
\\
4. STM images of \WO3/a-\TiO2(101) at higher coverage\\
\\
5. Distribution analysis of the M features\\
\\
6. \WO3 configurations calculated with DFT\\
\\
7. Schematic models of the M, C and 3M features\\
\\
8. \WO3 powder deposition onto r-\TiO2(110)

\newpage

\section{a-\TiO2(101) surface preparation}

Since large size anatase \TiO2 single crystals are difficult to synthesize in the lab,  mineral anatase \TiO2 single crystals (SurfaceNet GmbH, Germany) are used in this study. The crystals are cut into a $4\times4\times1$ (mm) dimension and are mounted on Ta sample plates clamped by Ta and Au strips. Due to the natural origin, the mineral anatase \TiO2 crystals contain certain amount of contaminates such as K, Fe, Nb, etc \cite{RN419}. Newly received a-\TiO2(101) crystals are subjected to at least 10 cycles of sputtering/annealing before STM images of good quality can be obtained. A typical cleaning cycle involves  Ar$ ^+$  (99.9995\%, Linde) sputtering (1.3 keV, 6-15 minutes) and UHV annealing (580-610 $^{\circ}$C, 20-40 minutes). The Ar$ ^+$ sputtering position is carefully calibrated to avoid sputtering contaminants onto the sample. Annealing in O$_2$ is found to be helpful in extracting contaminates and is performed occasionally to deplete contaminations in the near surface region. The annealing temperature is increased about 2 degrees per 10 cycles to facilitate the formation of  large terraces. Gradually, the crystal becomes more reduced and the color of the sample changes from a semi-transparent orange color into non-transparent black color. Throughout this work, three a-\TiO2(101) single crystals have been used. The STM images are collected on 2 different samples and the XPS data is obtained from the third crystal. No obvious changes in the morphology of supported \WO3 were observed on the two a-\TiO2 crystals. Thus, the influence of bulk reduction on the \WO3 structure, if any, is considered to be negligible. 
\section{\WO3 deposition methods}

The \WO3 powder (Sigma-Aldrich,  99.995\%) is contained in an \Al2O3 liner which is inserted in a Ta crucible mounted in a 4-pockets e-beam evaporator (EBE-4, Oxford Applied Research, UK). The deposition rate of \WO3 from the powder source is about 0.03 ML/min, as estimated by STM. For the reactive deposition of \WO3, a 2.0 mm W metal rod (99.99+\%, Goodfellow, UK) with sharpened tip is used. During W evaporation, the chamber pressure remains in the low $10^{-9}$ mbar range. O$_2$ (99.9995\%, Linde) is backfilled into the chamber to maintain a partial pressure of $1.0\times10^{-6}$ mbar during the growth of \WO3. The reactive deposition rate of \WO3 is calibrated to be around 0.06 ML/min. After the deposition of \WO3, the sample is usually annealed at 500 K for 5 minutes to facilitate easier STM scanning.

\section{XPS of \WO3/a-\TiO2(101)}

\renewcommand{\thefigure}{S1}
\begin{figure}[t]
\begin{center}
\includegraphics[width=0.8\textwidth]{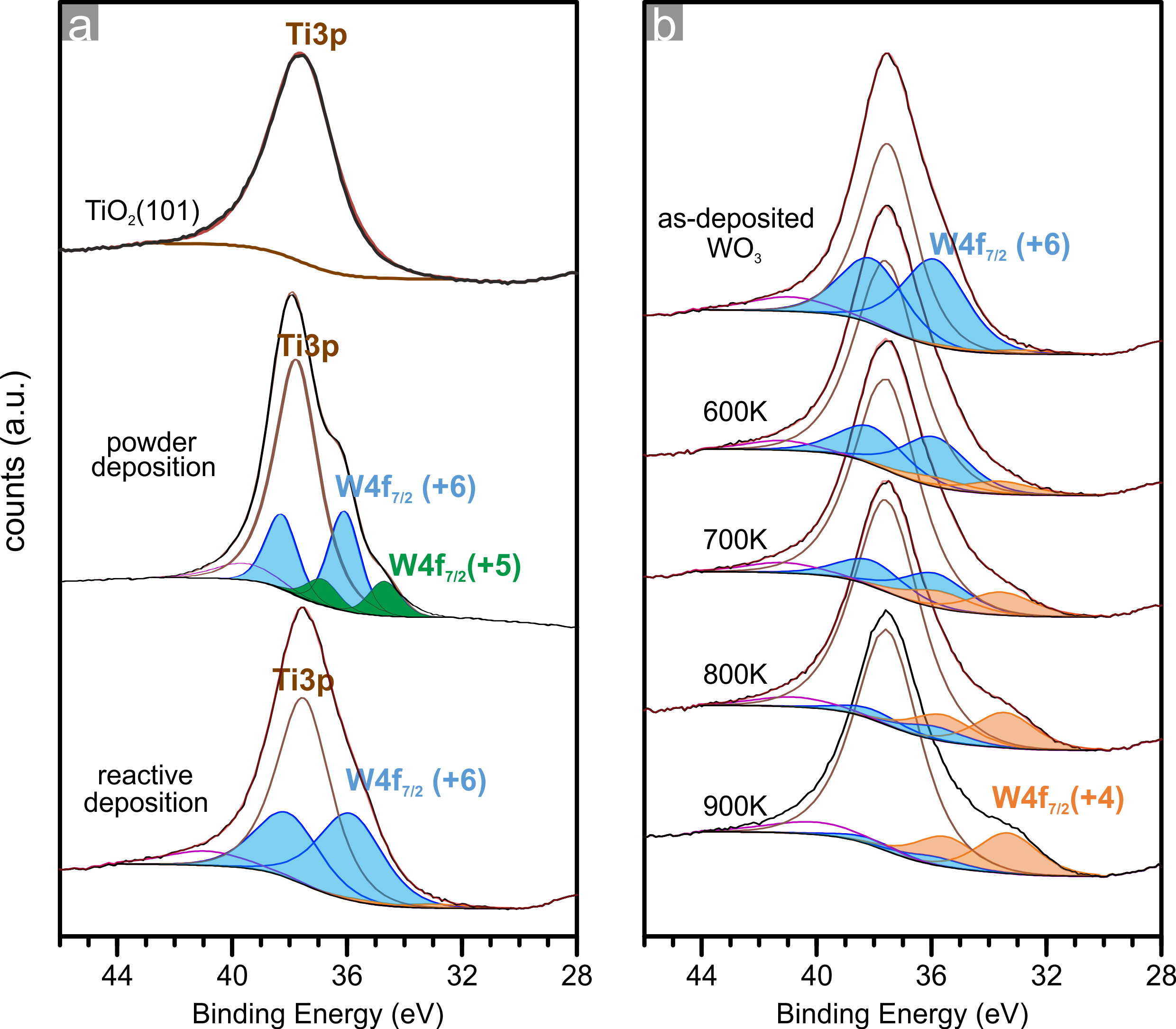}
\end{center}
\caption{(a) XPS spectra of the W4f and Ti3p regions for the reactive deposited and powder sublimated \WO3 species on a-\TiO2(101). (b) XPS spectra (W4f and Ti3p) obtained after annealing of the reactive deposited \WO3 at various temperatures. Spectra of powder deposited \WO3 were collected with a photon energy of 156.0 eV at the MalLine.}
\label{XPS}
\end{figure}

XPS spectra of reactive deposited \WO3 were collected at a commercial chamber (SPECS) with Al anode as the X-ray source (non-monochromatized) while the powder deposited \WO3 were collected at the ASTRID2 synchrotron (MatLine) facility at Aarhus University. The Matline is equipped with an SX-700 monochromator and a SPECS Phoibos 150 electron energy analyzer, which was operated at a pass energy of 20 eV and a curved analyzer slit of 0.8 mm \cite{RN1484}. The beamline monochromator exit slit was set to a width of 30 $\mu$m. The base pressure in the end-station was $3.0\times10^{-10}$ mbar. The O 1s core levels were collected with a photon energy of 651.15 eV and the W4f and Ti3p core levels were collected using 156.0 eV photon energy.

XPS measurements were performed to determine the oxidation state of the \WO3 species deposited on the a-\TiO2(101) surface.  Figure \ref{XPS}(a) shows the W4f/Ti3p regions of a clean a-\TiO2(101) surface and two differently prepared \WO3/\TiO2(101) samples. Due to the  overlapping of the W4f and Ti3p peaks, a careful deconvolution process is needed to extract the W4f signals. We found that the majority of  the as-deposited \WO3 species from the powder source are in the highest W oxidation state of +6 (W4f$_{7/2}$ at 35.7 eV) with a small portion of W in the +5 state ($<$ 25\%, W4f$_{7/2}$ at 34.5 eV). W in the +5 oxidation state originates from the bulk reduced WO${_{3-x}}$ powder in the evaporator and is not a general feature of deposited tungsta on a-\TiO2(101).The reactively deposited tungsta species are nearly completely in the +6 oxidation state (W4f$_{7/2}$ at 35.7 eV). In addition, we found by XPS that W(+6) state is thermally stable until 600 K, where it starts to be reduced to W(+4) (W4f$_{7/2}$ at 33.7 eV ), as shown in Figure \ref{XPS}(b). Above 600 K, a quick reduction happens. Following annealing at 800 K , the W(+6) is nearly completely converted to W(+4). Although the W oxidation state changes upon annealing at this high temperature, it is worth noticing that the total W4f peak areas decreases only slightly after annealing, indicating that W is thermally stable on the surface without much loss. The XPS results confirm that annealing the as-deposited \WO3/a-\TiO2(101) samples at 500 K does not reduce the deposited tungsta. Accordingly, brief annealing at 500 K, as done in the STM experiments, does not change the W oxidation state.

A Shirley background subtraction was used before fitting the W4f/Ti3p  XPS regions. The XPS spectrum of the clean a-\TiO2(101) surface provides a reference for  the calibration of Ti3p peak shape and position. The W4f$_{7/2}$ and  W4f$_{5/2}$ spin-orbital splitting peaks are constrained to be separated by 2.18 eV and  the peaks have identical FWHM with an area ratios of 4 to 3. For the spectra obtained with the lab X-ray source, the W4f peaks are fitted with 100\% Gaussian, while the Ti3p peak is fitted with 72\% Lorentzian + 28\% Gaussian. The XPS spectra collected at the synchrotron are characterized by more narrow peaks (smaller FWHM) and the shape of the spectra are adjusted by mixing 50\% Gaussian with 50\% Lorentzian. The W(+6) oxidation state is calibrated by measuring the W4f peaks from a \WO3 multilayer, deposited from the powder source on a-\TiO2(101) (data not shown). A shoulder peak at around 42.0 eV is also fitted which is assigned to the loss feature of \WO3.

Additionally, we utilized XPS to check the cleanliness of the bare and the tungsta-covered a-\TiO2(101) surfaces. Since the carbon amounts were negligible following \WO3 deposition from the powder source and reactive deposition, both preparation methods were accomplished in a very clean manner.

\section{STM images of \WO3/a-\TiO2(101) at higher coverage}

\renewcommand{\thefigure}{S2}
\begin{figure}
\begin{center}
\includegraphics[width=\textwidth]{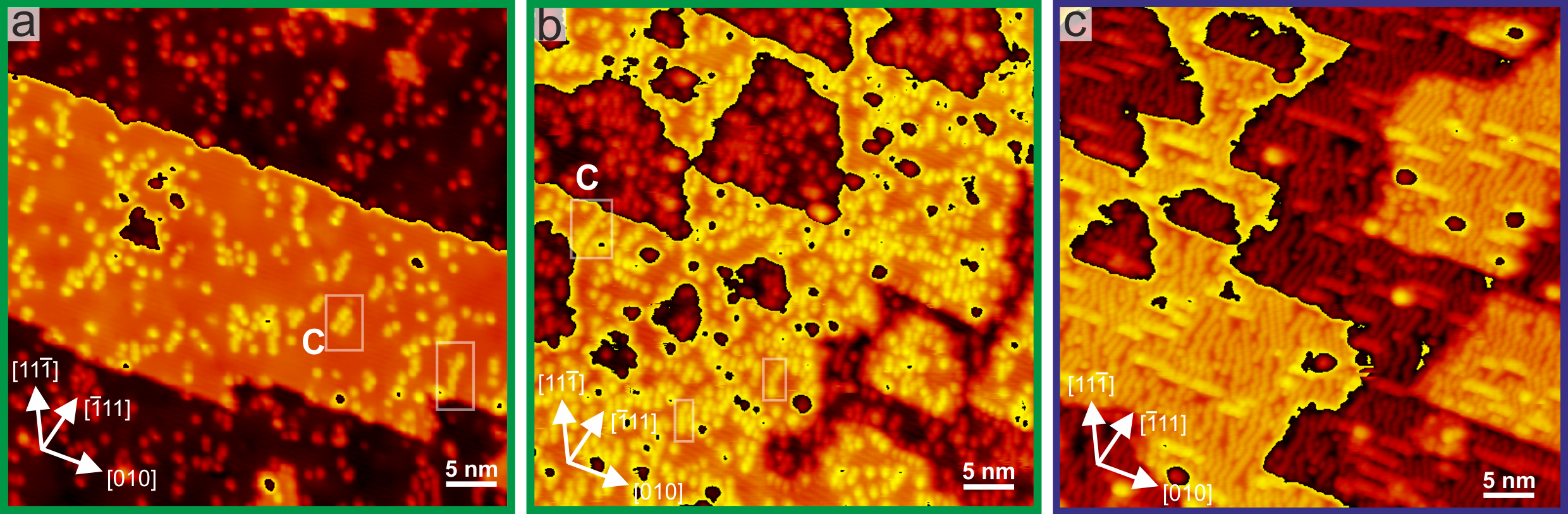}
\end{center}
\caption{STM images (500 \AA $\times$ 500 \AA) of reactive deposited \WO3 on a-\TiO2(101) at (a) intermediate and (b) high coverage in the submonolayer range. (c) STM of powder sublimated \WO3 on a-\TiO2(101) at full monolayer coverage.}
\label{reac_high}
\end{figure}

The STM scans were performed in a home-built Aarhus STM which is mounted in an UHV chamber with base pressure of $8.0\times10^{-11}$ mbar. The STM images were obtained with the sample at room temperature. A typical scanning condition is: 1.0-1.2 V (tip), 0.1-0.15  nA. 

Figure \ref{reac_high}(a-b) show the STM images of reactive deposited \WO3 on the a-\TiO2(101) surface at intermediate (Figure \ref{reac_high}(a)) and high coverage (Figure \ref{reac_high}(b)) after annealing at 500 K for 5 minutes. As the coverage increases, the M features appear closer to each other and eventually form chain-like C features along the [$\bar{1}11$] and [$11\bar{1}$] direction. Figure \ref{reac_high}(c) shows a full monolayer of powder sublimated \WO3 on the a-\TiO2(101) after annealing at 500 K for 5 minutes. The a-\TiO2(101) is nearly completely covered by the C features, forming small sizes of ($2\times1$) superstructure domains that are oriented along the [$\bar{1}11$] or [$11\bar{1}$] direction. 
\section{Distribution analysis of the M features}

\renewcommand{\thefigure}{S3}
\begin{figure}[t]
\begin{center}
\includegraphics[width=\textwidth]{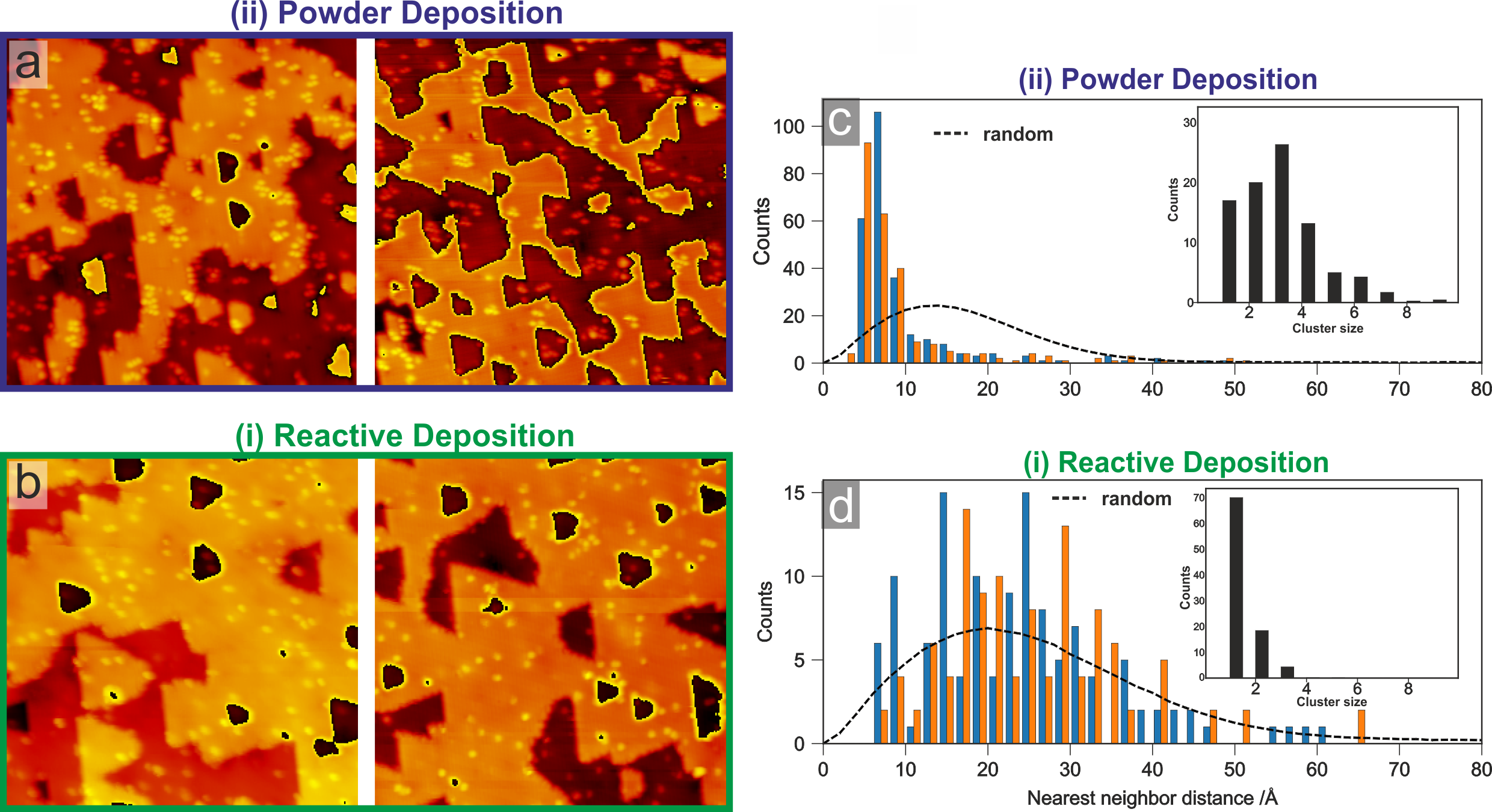}
\end{center}
\caption{STM images of (a) powder sublimated and (b) reactive deposited submonolayer \WO3  on a-\TiO2(101) used for the statistic analysis (500 \AA $\times$ 500 \AA). The samples are annealed at 500 K for 5 minutes before taking STM images at RT. Histogram of the NND distribution of the (c) powder sublimated and (d) reactive deposited \WO3 species on a-\TiO2(101).  The simulated random distribution is depicted as dashed line. Insets in (c) and (d) show the cluster size distribution determined by counting the number of monomers within a circle with a radius of 19.5 \AA.}
\label{histo}
\end{figure}

To quantitatively describe the distribution of the M features at low coverage, a stochastic analysis is performed by counting the nearest-neighbor distance (NND) distribution in the STM images, as shown in Figure \ref{histo}. The number of M features are counted directly from STM images in Figure \ref{histo}(a, b). The NND for random distribution is performed on a 500 \AA $\times$ 500 \AA 2-D space with the same coverages as found experimentally corresponding to Figure \ref{histo}(a, b). Note that the a-\TiO2(101) surface lattice was not imposed in the simulation. In total, 6000 iterations were performed to achieve convergence of the NND random distribution curve, as shown in Figure \ref{histo}(c, d). Clearly, the powder deposited \WO3 species are characterized by small NND that peaks between 4 - 8 \AA\ (note that the smallest possible NND of the M features is 7.58 \AA\ along the [010] direction and 5.46 \AA\ along the [$\bar{1}11$] and [$11\bar{1}$] directions) which is at a smaller distance than found in the simulated random distribution curve. The reactively deposited \WO3 species, however,  are characterized by a wider range of NND that peaks at around 20 \AA, which fits well to the simulated random distribution. Further more, it is clear that the powder deposited M features have a tendency to form clusters of trimers while the reactively deposited M features appear as monomers, as can be seen in the inset of Figure \ref{histo}(c, d). The cluster size is determined by searching for the number of M features within a circle with a radius of 19.5 \AA (this radius value is chosen to be larger than the length of a 3M chain (15.16 \AA) but smaller than a 4M chain (22.74 \AA) along the [010] direction).

\section{\WO3 configurations calculated with DFT}

\renewcommand{\thefigure}{S4}

\begin{figure}
\begin{center}
\includegraphics[width=0.8\textwidth]{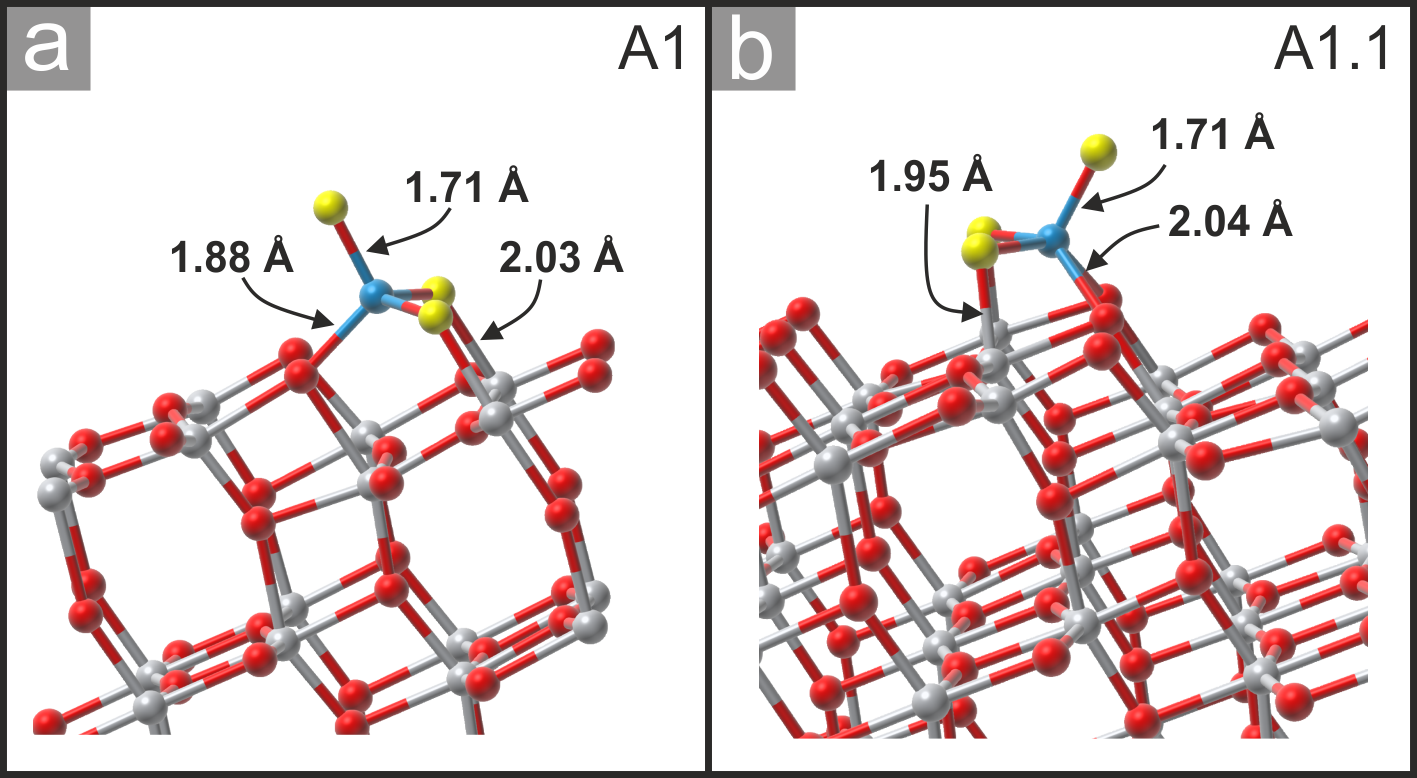}
\caption{DFT calculated \WO3 monomer structures on a-\TiO2(101): (a) the most stable \WO3 monomer, A1. (b) the second most stable \WO3 monomer, A1.1. Gray balls: Ti atoms; red balls: O atoms; yellow balls: O atoms in \WO3; blue balls: W atoms in \WO3.}
\label{monomer}
\end{center}
\end{figure}

\renewcommand{\thefigure}{S5}
\begin{figure}
\begin{center}
\includegraphics[width=0.8\textwidth]{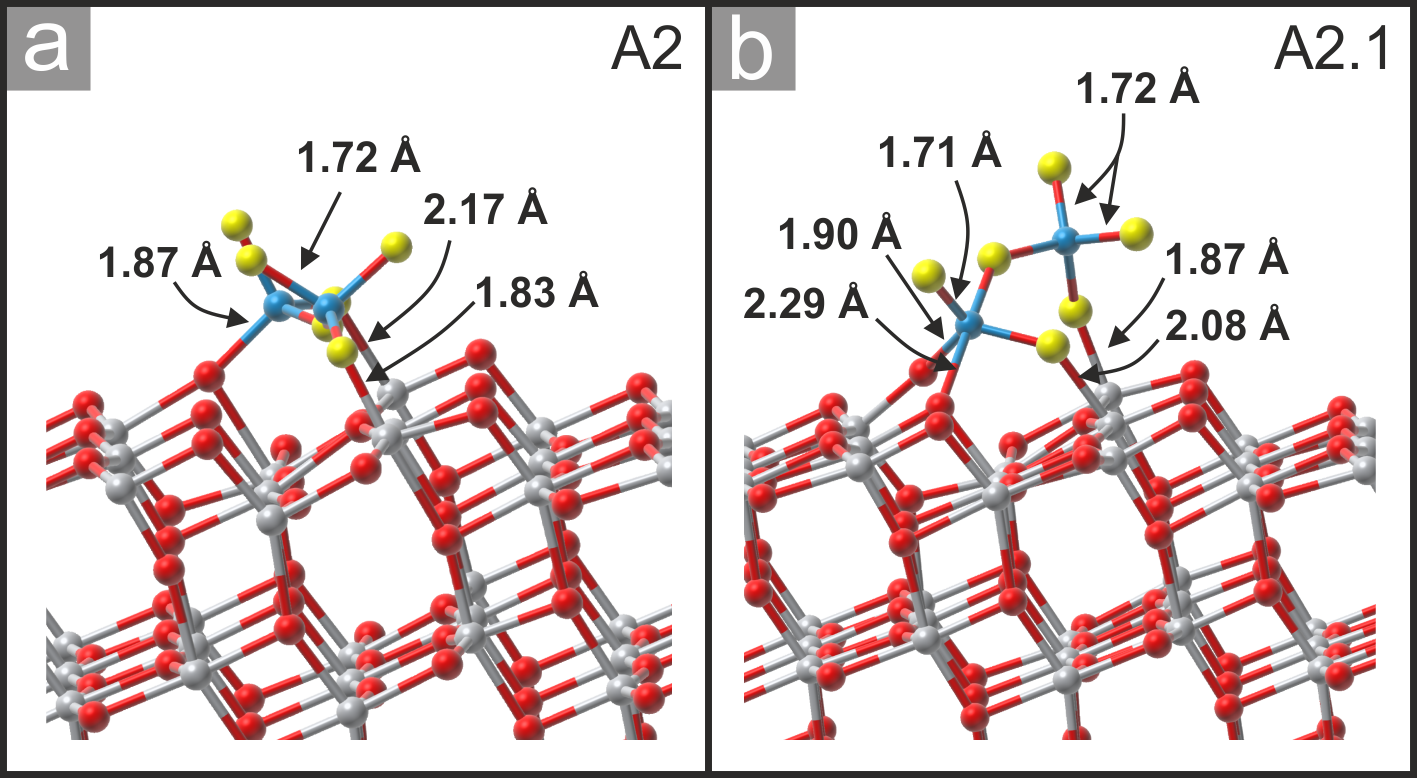}
\caption{DFT calculated W$_2$O$_6$ dimer structure on a-\TiO2(101): (a) the most stable W$_2$O$_6$ dimer, A2. (b) the second most stable W$_2$O$_6$ dimer, A2.1. Ti atoms; red balls: O atoms; yellow balls: O atoms in \WO3; blue balls: W atoms in \WO3. }
\label{dimer}
\end{center}
\end{figure}

\renewcommand{\thefigure}{S6}
\begin{figure}
\begin{center}
\includegraphics[width=0.8\textwidth]{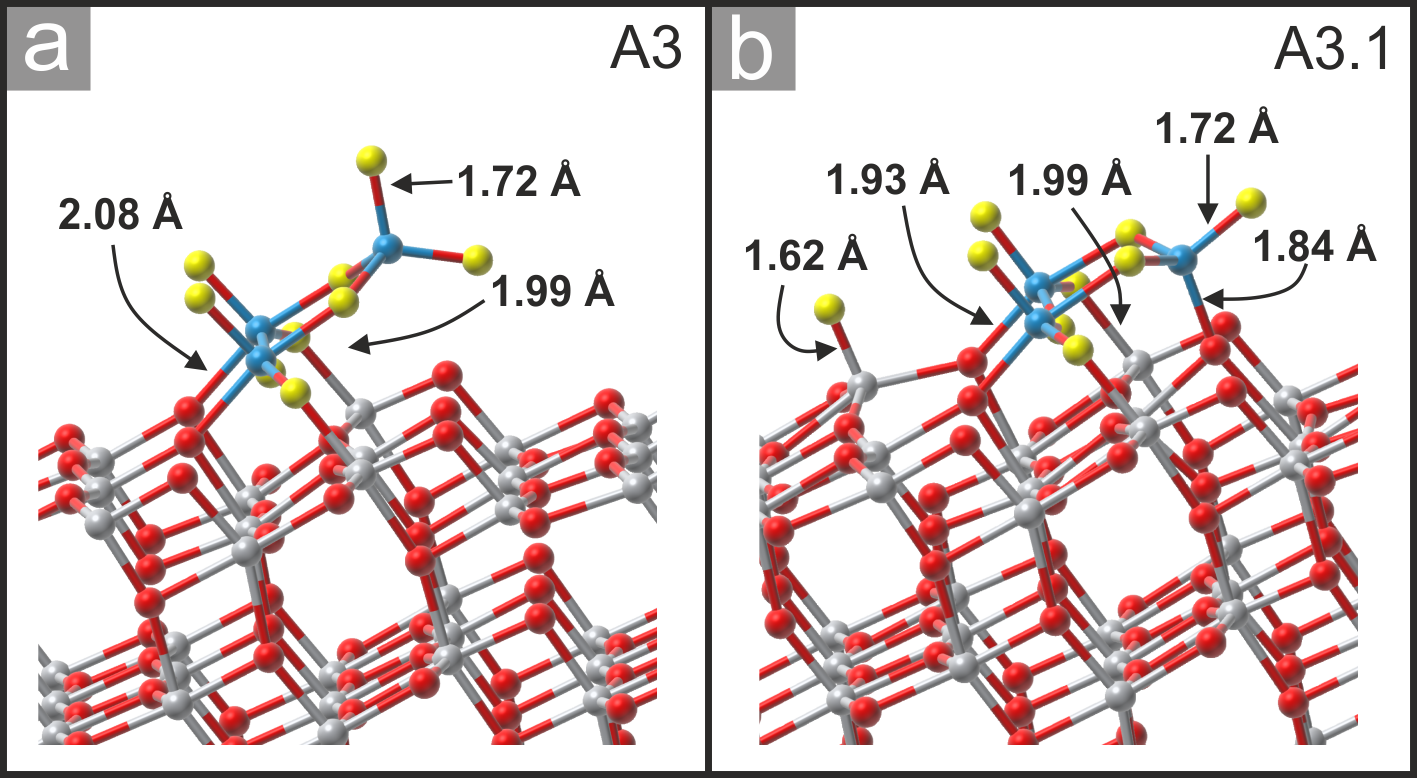}
\caption{(a) DFT calculated  W$_3$O$_9$ trimer structure on a-\TiO2(101), A3. (b) DFT calculated W$_3$O$_8$* + O* configuration on a-\TiO2(101), A3.1. Configuration A3.1 is more stable than A3. Ti atoms; red balls: O atoms; yellow balls: O atoms in \WO3; blue balls: W atoms in \WO3.}
\label{trimer}
\end{center}
\end{figure}

Density Functional Theory (DFT) calculations are performed using the Grid-based Projector Augmented Wave (GPAW) package \cite{RN1485,RN1486} through the ASE package \cite{RN1487}. The crystal structure of \WO3 was optimized using the stress tensor method available in ASE to be a=7.373 \AA, b=7.650 \AA\ and c=7.819 \AA, which deviates from the experimental value with less than 2.5\% \cite{RN1490}. The a-\TiO2(101) surface was modelled with a super cell size of (3$\times$1) and a slab of 3 layers of \TiO2, with the bottom layer atoms kept fixed at their bulk positions. The cells were periodic in the \textit{xy} plane with a minimum of 10 \AA\ between slabs in the \textit{z}-direction, and the lattice constants  taken from reference \cite{RN443}.  To obtain the most stable structures of the (\WO3)$_{x}$ species on the a-\TiO2(101) surface we used a genetic algorithm (GA) \cite{RN1488}. A two-step optimization scheme was applied to explore a large area of the potential energy surface (PES). The GPAW program supports three different operating modes for running DFT calculations: The LCAO mode where the wave functions are represented with an atomic orbital basis set, the PW mode, where the wave functions are expanded as plane waves, and grid-based mode, where on a uniform real space orthorhombic grids. Since the LCAO mode with dzp basis set is fastest, we first used this to scan the PES with the GA. To find the most stable structure we then optimized the structures within 1 eV of the most stable structure found with the GA in the more accurate grid-based mode with h= 0.18, and within the generalized gradient approximation with the BEEF‐vdW functional \cite{RN1489}.  The DFT energies of the most and second most stable monomeric, dimeric and trimeric species are given in Table \ref{tab-SI}. 

To compare the relative stability of monomeric, dimeric and trimeric species, we have calculated their stabilities per formula \WO3 unit. As reference, we take the total energy of the most stable adsorbed \WO3* monomer (A1):  $$E{_{W{_x}O{_{3x}}, stability}} = (E{_{W{_x}O{_{3x}}}} – xE{_{WO{_3}{^*} [A1]}} + (x-1)E{_{a\mh TiO{_2}(101)}})/x $$ A positive $E{_{W{_x}O{_{3x}}, stability}} $ indicates that the considered species is less stable than the most stable monomeric \WO3 on a-\TiO2(101).

\renewcommand{\thetable}{S1}
\begin{table}
  \begin{center}
    \caption{Total energies are DFT calculated energies}
    \label{tab-SI}
    \begin{tabular}{l|r|r} 
      \textbf{Species} & \textbf{Energy (eV)}&\textbf{\textit{E}}${_{W{_x}O{_{3x}}, stability}}$ \textbf{(eV)}\\
      \hline
      a-\TiO2(101)& -5255.57 & \\
      \hline
    \WO3*(A1)& -5622.91& 0\\
      \hline
      \WO3*(A1.1)&-5622.21& 0.700\\
      \hline
     W$_2$O$_6$*(A2)&-5989.31&0.470\\
      \hline
      W$_2$O$_6$*(A2.1)&-5989.20&0.525\\
      \hline
      W$_3$O$_9$*(A3)&-6356.34&0.417\\
      \hline
     W$_3$O$_8$* + O*(A3.1)&-6356.48&0.370\\

    \end{tabular}
  \end{center}
\end{table}

The A1 configuration (\ref{monomer}(a)) is a \WO3 monomer with three bonds to the a-\TiO2(101) surface, including two WO-Ti(5f) bonds (2.03 \AA) and one W-O(2f) bond (1.88 \AA). The terminal W-O bond length (1.71 \AA) is small, suggesting a double bond. The W cation is in a tetrahedral environment surrounded by four O ions. Figure \ref{monomer}(b) shows the second most stable monomer structure that is labelled with A1.1. In this structure, the \WO3 adsorbate has four bonds to the a-\TiO2(101) surface, including two WO-Ti(5f) (1.95 \AA) and two W-O(2f) (2.04 \AA) bonds. The A1.1 configuration is 0.7 eV less stable than the A1 configuration.

The A2 configuration (Figure \ref{dimer}(a)) - the most stable W$_{2}$O$_{6}$ dimer - is bound via three bonds to the a-\TiO2(101) surface, including two WO-Ti(5f) bonds (1.83 \AA, 2.17 \AA) and one W-O(2f) bond (1.87 \AA). The dimer spans across three Ti(5f) sites along the [010] direction. The second most stable W${_2}$O${_6}$ dimer - labelled A2.1 - is shown in Figure \ref{dimer}(b). In this structure, the dimer is bound by four bonds to the a-\TiO2(101) surface, including two  WO-Ti(5f) bonds and two W-O(2f) bonds. This configuration is 0.11 eV less stable than the A2 structure.

The intact W$_3$O$_9$ trimer, configuration A3 (see Figure \ref{trimer}(a)), is characterized by four bonds to the a-\TiO2(101)  surface, including two WO-Ti(5f) bonds (1.99 \AA) and two W-O(2f) bonds (2.08 \AA). The molecular plane of the cyclic W$_3$O$_9$ trimer is tilted away from the surface. The configuration labelled A3.1  (see Figure \ref{trimer}(b)) consists of a W$_3$O$_8$ species and a nearby terminal O atom, bound to a Ti(5f) site. The bond length of the terminal O atom is only 1.62 \AA, indicating a double bond. The W$_3$O$_8$ species binds to the a-\TiO2(101) with five bonds, including two WO-Ti(5f) bonds (1.99 \AA) and three W-O(2f) bonds (1.84 \AA, 1.93 \AA, 1.96 \AA). Notice that the A3.1 configuration is more stable than the A3 structure by 0.14 eV.

\section{Schematic models of the M, C and 3M features}

On the basis of the A1 structure (\WO3 monomer) found with DFT calculations, the M, C and 3M features can be easily constructed, as shown in Figure \ref{Scheme}(a) shows how the C features can be composed. Figure \ref{Scheme}(b) shows the different kinds of 3M features, which we indeed observed in our STM studies.

\renewcommand{\thefigure}{S7}
\begin{figure}
\begin{center}
\includegraphics[width=0.8\textwidth]{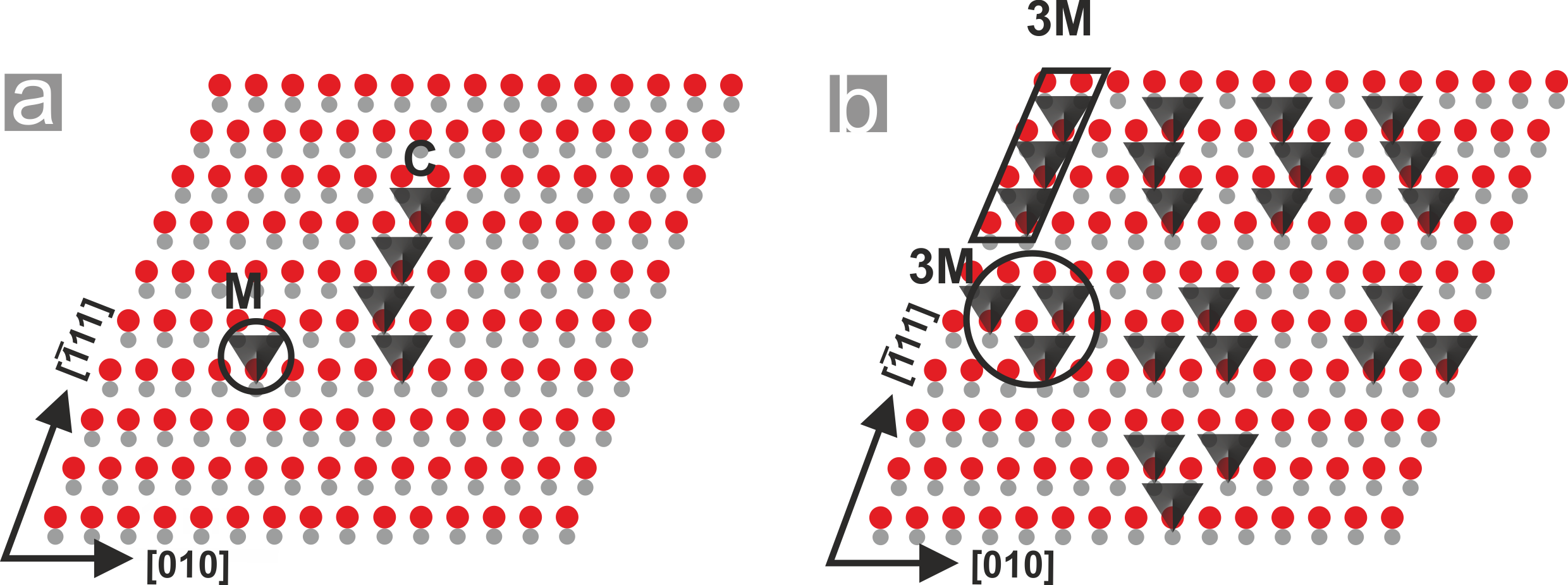}
\end{center}
\caption{Schematic models of the (a) M, C and (b) 3M features on a-\TiO2(101) constructed from the \WO3 monomer model (A1 configuration). (black triangle: \WO3 monomer; gray balls: Ti(5f) atoms; red balls: O(2f) atoms.) }
\label{Scheme}
\end{figure}

\section{\WO3 powder deposition onto r-\TiO2(110)}

\renewcommand{\thefigure}{S8}
\begin{figure}[t]
\begin{center}
\includegraphics[width=\textwidth]{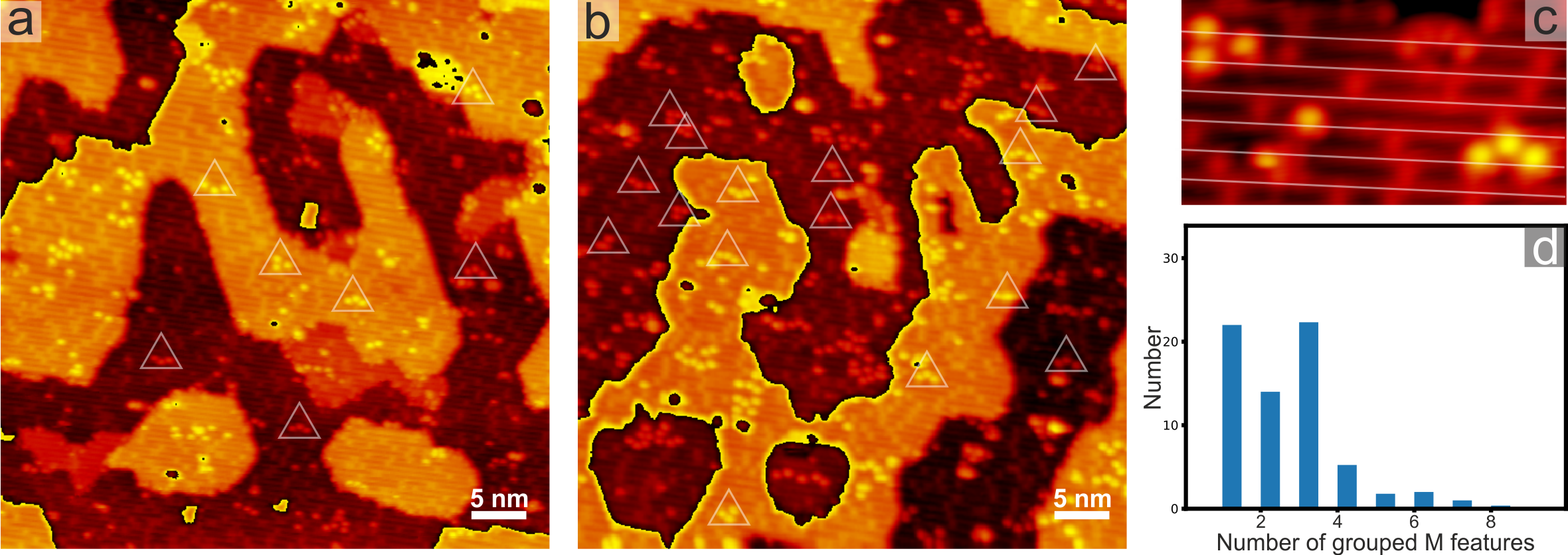}
\end{center}
\caption{(a, b) STM images (500 \AA $\times$ 500 \AA) of the powder sublimated \WO3/r-\TiO2(110) samples after brief annealing at 500 K. The \WO3 coverage in (b) is higher than that in (a). White triangles indicate the 3M features. (c) Zoom-in STM image (97 \AA $\times$ 51 \AA) with 3M features. (d) Cluster size distribution of the tungsta species seen in (a). The cluster size was determined by counting the number of tungsta species in a circle with a radius of 20 \AA. This radius was chosen to accommodate three tungsta species (M features) along the [001] direction.}
\label{rutile}
\end{figure}

In addition to \WO3/a-\TiO2(101), we also utilized STM to study \WO3/r-\TiO2(110). These STM studies were conducted to address the apparent disagreement regarding the assignment of tungsta-related species on \TiO2 surfaces, a-\TiO2(101) and r-\TiO2(110), respectively. Or is the interaction of tungsta with the a-\TiO2(101) surface very different from the tungsta–r-\TiO2(110) interaction? We used \WO3 powder sublimation method to prepare \WO3/r-\TiO2(110) sample, exactly as in previous studies by Dohnálek and co-workers \cite{RN1250,RN1492}. Upon \WO3 powder sublimation, the r-\TiO2(110) surface was at RT. Subsequently, the \WO3/r-\TiO2(110) sample was vacuum-annealed at 500 K for 5 minutes. 

As shown in Figure \ref{rutile}(a, b), we successfully prepared \WO3/r-\TiO2(110) samples characterized by a rather low tungsta coverage. Tungsta-related new protrusions on r-\TiO2(110) appear with lateral sizes of $ \sim$6.2 \AA and STM heights of $\sim$1.5 \AA. These dimensions are identical to those found for the M features on a-\TiO2(101). In addition, the new protrusions on r-\TiO2(110) appear preferentially in groups of three, even though some isolated protrusions also occur. Examples of grouped tungsta-related protrusions are highlighted by white triangles in Figure \ref{rutile}(a, b). That numerous groups of three protrusions occur is likewise as found for \WO3/a-\TiO2(101) samples prepared by \WO3 powder sublimation. Accordingly, the STM data presented in Figure \ref{rutile}(a–c) indicate that the same type of tungsta species are formed on r-\TiO2(110) surfaces as on a-\TiO2(101) surfaces (i.e., the M features). Thus, the interaction of (\WO3)$_3$ species with r-\TiO2(110) is very similar to the (\WO3)$_3$–a-\TiO2(101) interaction. As argued in our main manuscript addressing \WO3/a-\TiO2(101), we assign the M features to \WO3 monomers.

On the one hand, the STM data presented here are consistent with our interpretation of dissociated (\WO3)$_3$ species on a-\TiO2(101). On the other hand, these STM data are in contrast to the assignment put forward in the studies by Dohnálek and co-workers addressing \WO3/r-\TiO2(110) \cite{RN1250,RN1492}, where the smallest tungsta species on r-\TiO2(110) were assigned to cyclic (\WO3)$_3$ trimers. Notice that the size of the smallest tungsta species are very similar in the STM studies of both research groups (ours and Dohnálek and coworkers).

Because of the low coverage on our \WO3/r-\TiO2(110) samples, it was possible to conduct a statistical analysis of the local clustering effect of the tungsta species, see Figure \ref{rutile}(d). It can be seen that there are peaks for isolated M features and 3M features.  The fact that tungsta species appear in groups of three as observed in our studies for \WO3/a-\TiO2(101) and \WO3/r-\TiO2(110) cannot be easily explained if the M features would originate from (\WO3)$_3$ species. However, such groups of three tungsta species can be explained straightforwardly within the model presented in our main manuscript. Within this model, (\WO3)$_3$ species impinge the \TiO2 surfaces, which then dissociate and slightly separate on the surfaces. In this light, the STM results obtained for \WO3/r-\TiO2(110) are further support for our assignment of the M features to \WO3 monomers. Moreover, it appears that (\WO3)$_3$ species dissociate on \TiO2 is not limited to the a-\TiO2(101) surface. Instead, our STM studies point to (\WO3)$_3$ dissociation on a-\TiO2(101) and r-\TiO2(110). It is possible that (\WO3)$_3$ species generally dissociate on \TiO2 surfaces at RT, regardless of the \TiO2 polymorph and the selected face.

Generally, it is challenging to clarify as to why the studies conducted in different laboratories casually lead to different conclusions. Also in the current case, this is difficult, and there are eventually several points to be considered why the M features were assigned differently. Nevertheless, when comparing the STM images published in Refs. \cite{RN1250,RN1492} with ours, we noted that the tungsta coverage was clearly higher in the work by Dohnálek and co-workers. We speculate that the high tungsta coverages in Refs. \cite{RN1250,RN1492} have led to “crowded situations” on the \WO3/r-\TiO2(110) samples that hindered Dohnálek and co-workers to recognize the grouping of M features.

\bibliographystyle{unsrt}
\bibliography{210}

\end{document}